# A Language Model of Java Methods with Train/Test Deduplication

Chia-Yi Su
csu3@nd.edu
University of Notre Dame
Notre Dame, IN, USA

Aakash Bansal
abansal@nd.edu
University of Notre Dame
Notre Dame, IN, USA

Vijayanta Jain
vijayanta.jain@maine.edu
University of Maine
Orono, ME, USA

Sepideh Ghanavati
sepideh.ghanavati@maine.edu
University of Maine
Orono, ME, USA

Collin McMillan
cmc@nd.edu
University of Notre Dame
Notre Dame, IN, USA

## ABSTRACT

This tool demonstration presents a research toolkit for a language model of Java source code. The target audience includes researchers studying problems at the granularity level of subroutines, statements, or variables in Java. In contrast to many existing language models, we prioritize features for researchers including an open and easily-searchable training set, a held out test set with different levels of deduplication from the training set, infrastructure for deduplicating new examples, and an implementation platform suitable for execution on equipment accessible to a relatively modest budget. Our model is a GPT2-like architecture with 350m parameters. Our training set includes 52m Java methods (9b tokens) and 13m StackOverflow threads (10.5b tokens). To improve accessibility of research to more members of the community, we limit local resource requirements to GPUs with 16GB video memory. We provide a test set of held out Java methods that include descriptive comments, including the entire Java projects for those methods. We also provide deduplication tools using precomputed hash tables at various similarity thresholds to help researchers ensure that their own test examples are not in the training set. We make all our tools and data open source and available via Huggingface and Github.

## 1 INTRODUCTION

Large language models (LLMs) have quickly become central to many areas of research [38]. Within Software Engineering (SE), they now form the basis for many approaches to code completion [30], automatic documentation generation [1], automatic bug repair [21], and dialogue systems about code [32]. Meanwhile, research in several domains has led to LLMs with ever-increasing parameter counts and training data. These LLMs have shown remarkable performance on several natural language problems, with products such as OpenAI's ChatGPT [28] and GitHub's Copilot [12] reaching stratospheric levels of public attention. Application of this technology to SE has naturally caught the eye of many researchers.

The typical application of LLMs in research is to use a "foundation model" pretrained on big data, followed by fine-tuning steps to customize the LLM to a specific problem. In research, high value is placed on control of experimental variables, and care must be taken to avoid contaminating the training set with test data. SE researchers typically work at a relatively fine grain, such as subroutines, statements, or even individual variables, and Java is a very popular language to study. Thus, what many SE researchers need is an LLM pretrained on a known set of Java methods, that they can then fine-tune at low cost for various SE research tasks.

Yet a caveat for researchers is that many LLMs are closed-source, have opaque training data and procedures, and/or are far too large and complex to reproduce in a laboratory setting. Hellendoorn and Sawant [16] point out that the situation is becoming inaccessible to researchers, which has negative implications for scientific rigor as published results cannot be reproduced or closely scrutinized. A metaphor for the current situation is that SE researchers need a "mouse model" of LLMs for code. In biology, mice and other animals are used to test ideas in a laboratory before the ideas are ready to be scaled to humans. Likewise, we need inexpensive models that are nonetheless similar enough to industrial LLMs to suggest that ideas that work in the lab can work in practice.

In this tool demonstration, we release a 350m parameter GPT2-like language model in three variations: one pre-trained with a dataset of 52m Java methods, one pre-trained with a dataset of 13m StackOverflow threads, and one trained on both. We also provide: 1) A held-out set of 8,192 Java methods for testing various research applications. 2) Precomputed deduplication tool for researchers to easily check if their examples are in the training set, so they can avoid training set contamination. 3) Tools and instructions for fine-tuning the models on different SE research tasks.

## 2 BACKGROUND & RELATED WORK

A "language model" is a probability distribution given a sequence of natural language tokens. The tokens may be characters, words, or sub-word units. The list of all known tokens is called the "vocabulary." In practice, the input to a language model is a sequence of tokens, and the output is (usually) the probability that each word in the vocabulary will be the next token in the sequence. In recent years, language models based on neural models have proliferated. These models are given a set of "training data" from which to generate the probability distribution. The size of the neural models and training data have increased multiple orders of magnitude in a small number of years. Today, language models consisting of billions of parameters and trillions of tokens of training data are the heart of several high-profile products such as ChatGPT and Copilot [19].

The steps to use a language model in a product usually consist of: 1) obtain a "foundation model" that is pretrained on a very large dataset of general domain text, and 2) "fine tune" the model on problem-specific input/output examples. As a general rule, the bigger the better. State-of-the-art results are possible with foundation models in the tens or hundreds of billions of parameters, pretrained with datasets encompassing the entire internet, libraries of books, legal documents, etc. Fine-tuning these models becomes a resource problem, as even loading the models requires tens of gigabytes of memory, and computation requirements are high with even a few examples [7]. An alternative is a partial fine-tune procedure such as LoRA [17], which trades accuracy for reduced computational cost and yet more experimental variables (e.g., LoRA parameters) [9].

A plethora of LLMs have been released as potential foundation models recently, such as GPT-2/3 [8], LLaMA [36], GPT-J [40], and GPT-NeoX [6]. Many even target source code, such as CodeParrot [37], StarCoder [39], and CodeGeeX [41]. The training data for these consists of datasets such as The Stack [23], which contains 1.5TB of code in 317m files in over 350 programming languages. Researchers in several areas of software engineering have indicated how the "fine-tuned LLM" strategy is likely to achieve state-of-the-art results when foundation models are further trained with task-specific examples in problems such as code generation [34], code summarization [1], and clone detection [29].

The problem for researchers is that the results from many foundation model LLMs are difficult to understand from a scholarly point of view. While the results may be state-of-the-art, it is difficult to eliminate experimental variables such as architectural differences in the models, choices of hyperparameters, and sources of data contamination between training and test sets. If a researcher's test set contains e.g. 10,000 Java files, the cost to ensure deduplication between that test set and e.g. The Stack becomes 10k x 317m = 3.17T file comparisons. If that researcher also wishes to verify if a different set of hyperparameters during pretraining would lead to different results, the cost to retrain the LLM from scratch could be tens or hundreds of thousands of dollars [33]. At these sizes, even basic principles of scientific integrity are cost prohibitive.

The problem of LLM cost to scientific research is earning more attention. Hellendoorn and Sawant [16] crystalize several of the problems and propose a few solutions. In the community more broadly, open-source solutions are becoming popular, though many of these still contain noisy datasets or are difficult to reproduce.

## 3 THE TOOL

This section introduces our tool, including the datasets, language model, and supporting fine-tuning and deduplication toolkits.

### 3.1 Target Audience / Requirements

Our target audience includes researchers in SE who study Java, particularly at the level of methods, statements, and variable names. The key requirements of our tool are:

1. A training set at the level of Java methods that are searchable in tractable time for duplicates or other attributes.
2. An additional training set containing natural language, within the domain of software engineering.
3. A model design that is easy to modify and well-studied in existing literature – no "magic."
4. A model size suitable for full retraining and fine-tuning (i.e. without LoRA) on a single 16GB GPU.
5. Clear instructions for reproducibility by students or others learning the technology's fundamentals.

Note that we **do not** seek to achieve state-of-the-art results. Our goal **is not** to be the largest or most-capable LLM, and we **do not** intend for our tool to be used in commercial products. Our target audience is scientific researchers and educators seeking maximum control of experimental variables within an accessible limit of resource constraints. This audience is likely to rerun experiments many times, across many machines, encounter errors (e.g., student learners), and/or operate in a cost restricted environment. At present, a single 16GB GPU workstation with a recent architecture (e.g., NVidia Ampere [27]) is available in most markets for under US$1,500.

### 3.2 Language Model

We build our language model using the GPT-2 model design as presented by Karpathy in the NanoGPT implementation [22]. We configure the model with the following parameters:

| | | |
|---|---|---|
| $e$ | embedding dimensions | 1024 |
| $L$ | number of layers | 24 |
| $h$ | attention heads | 16 |
| $c$ | block size / context length | 256 |
| $b$ | batch size | 4 |
| $a$ | accumulation steps | 32 |
| $d$ | dropout | 0.20 |
| $r$ | learning rate | 3e-5 |
| $y$ | weight decay | 1e-1 |

Parameters $e$, $L$, and $h$ correspond to the GPT-2-medium size (350m parameters). This size is large enough to be likely to produce meaningful results in many situations (e.g., BERT [10] is 345m parameters), while small enough to fit into 16GB VRAM. The context length $c$ of 256 tokens covers over 95% of Java methods in our dataset, and yet at much lower computational cost than the default 1024 size. Since our target audience use Java methods, statements, and variables, the smaller context length is appropriate. The batch size $b$ allows the model to be reproduced within the 16GB limit, but the accumulation steps $a$ of 32 maintains a relatively high effective batch size of 4 x 32 = 128. The dropout, learning rate, and weight decay are defaults of NanoGPT.

### 3.3 Datasets

We release two datasets that we use to pretrain our model.

**jm52m** is a dataset of 52m Java methods created from 52k Java projects. The source code originated from the Merobase [20] and Sourcerer [26] data releases, supplemented by our own prior work in LeClair *et al.* [24]. It contains code uploaded to code repositories between 2008 and 2018. We then extracted every Java method from every file and project. We removed empty methods, methods from corrupt files, and methods with parsing errors.

**so13m** is a dataset containing 13m discussion threads from StackOverflow. The origin of the data is the StackExchange data dump [18] from between January 2014 and December 2022. The threads cover a multitude of topics. This dataset serves as a natural language and (often) accompanying code in the domain of software engineering. Its inclusion could help downstream tasks depending on generating or understanding natural language.

We use the GPT byte-pair encoder [11] to tokenize both datasets. We provide a SQL database dump (see jm52m.sql on our website, Section 5) for traceability of the methods to their files and projects. The following table shows the dataset sizes in different metrics:

|  | jm52m | so13m |
| --- | --- | --- |
| number of tokens | 8,752,695,577 | 10,495,518,108 |
| number of documents | 51,841,717 | 13,071,148 |
| number of files | 8,402,038 | n/a |
| number of projects | 52933 | n/a |
| megabytes after processing | 16,695 | 20,019 |

We create a **holdout set** of 8,192 Java methods from jm52m. These are the Java methods from the test set for comment generation research published by Bansal *et al.* [3] in the funcom-java-long dataset. We chose these because they were filtered for quality, contain header comments which may assist various areas of research, and are consistent with other work for easier reproducibility. We exclude the holdout methods from jm52m and use them as the basis for deduplication in Section 3.5.

### 3.4 Model Releases

We call our model **jam** (for Java Methods). We release three versions of the model:

**jam** (default, also jam-jm) This model is trained on jm52m only. We train for one epoch, which is ~300,000 iterations. We intend this version for most applications.

**jam-so** This model is trained on so13m only. Also trained for one epoch, also about 300,000 iterations.

**jam-sojm** This model is trained on so13m and then jm52m for one epoch each after resetting the learning rate and decay.

Our training hardware consists of an Intel i9-10900X CPU with 128GB memory and 2xA5000 NVidia GPUs. Training time is approximately 2.1s per iteration, or about seven days per epoch. Scaling with faster hardware or more GPUs would likely accelerate training. The model files contain the iteration and training configuration. We provide instructions on our website for resuming training if more epochs are desired.

### 3.5 Deduplication Toolkit

We provide a deduplication toolkit to help researchers verify that their test sets are not in the training set. This toolkit is necessary because different users may have different tolerances for duplicates, and because not all users will want to use the holdout set we provide. For example, code generation experiments will have a low tolerance for duplicates since the model may have seen the code it is trying to generate. But some experiments, e.g. generating code embeddings, may have a higher tolerance since the model will see the code to generate the embedding vector anyway.

Our toolkit is based on the MinHashLSH deduplication technique [25], which is widely used in machine learning for deduplication [4, 5, 35]. The way MinHashLSH works is to generate an lsh object for a set of documents at a given threshold of similarity. Then, an alternate document is hashed and checked against that object. The result is a set of documents in the lsh object that the alternate document matches. Creating the lsh objects needs non-trivial computing power: to generate the objects for jm52m took about 26 hours on our workstation.

We provide the following:

1. A set of scripts using MinHashLSH to generate lsh deduplication objects from jm52m and so13m at a given threshold.
2. Precomputed lsh objects at four thresholds for each dataset.
3. Lists of Java methods from the holdout set which researchers may consider removing, depending on their tolerance for duplicates. We also provide the ID numbers of documents in jm52m and so13m that are considered matches at different similarity thresholds, to allow for manual inspection.
4. A program for using the precomputed lsh objects on given code, to quickly check for duplicates locally.
5. A web API and interface for checking a document online.

We chose the following thresholds for lsh objects in each dataset. The count indicates the number of methods in the holdout set that were detected as potential duplicates at that threshold. As a general rule, we found the most lenient options to include very near exact duplicates only, while the most strict options were quite far and may have only a few overlapping words.

| jm52m | | so13m | | |
| --- | --- | --- | --- | --- |
| threshold | count | threshold | count | |
| 0.5 | 4822 | 0.3 | 4979 | ↑ more strict |
| 0.6 | 1905 | 0.4 | 1486 | |
| 0.7 | 611 | 0.5 | 136 | |
| 0.8 | 44 | 0.6 | 0 | ↓ more lenient |

### 3.6 Fine-tuning Toolkit

We provide a fine-tuning toolkit to help researchers adapt jam to a specific research problem. Our toolkit is drawn from the NanoGPT framework, with a few customizations to streamline the fine-tuning process using our model. We integrate code by Grittner [13] to allow for fine-tuning using LoRA, in case researchers have even more limited resources than expected or want to answer research questions about the effects of LoRA on their particular problem. Our toolkit will automatically download our trained models from our Huggingface model repository, to minimize steps needed by the researcher. All code is MIT licensed.

## 4 APPLICATION

In this section, we demonstrate one application of our tool for the problem of code summarization, which is the task of writing comments that describe code [14]. Code summarization is an active research area that is suitable for the application of pretrained language models. A typical target for this problem in Java are the short header comments known as JavaDocs. Most approaches use an encoder-decoder neural architecture in which the source code is encoded and the descriptive comment is "decoded." These approaches work well considering limited data, though could be improved using larger models pretrained with relevant data.

To demonstrate how jam can help this line of research, we use the funcom-java-long dataset presented by Bansal *et al.* [3] from which we extracted the holdout set in Section 3.3. The training set for funcom-java-long has ~170k Java methods and summaries. These are already heavily cleaned and deduplicated from the holdout set by Bansal *et al.* [2]. From these 170k samples, we create fine-tuning example prompts in the form:

TDAT: <method code>   COMMENT: <comment> <!endofdoc>

Then we fine-tune for four epochs using a fixed (non-decaying) learning rate of 3e-5, with all other parameters equal to the table in Section 3.2. Our configuration information for this small experiment is in the file config/finetune_funcom.py on our website. We fine-tune our three models: jam, jam-so, jam-sojm. We also fine-tune pre-trained gpt2-medium [31] and an equally-sized NanoGPT model from scratch as a comparison points. All models have the same architecture and parameters, to simplify comparison.

|  |  | METEOR | USE | BLEU |
|---|---|---|---|---|
| full | jam | 33.25 | 51.29 | 20.07 |
|  | jam-so | 34.04 | **52.88** | 19.83 |
|  | jam-sojm | **34.61** | 52.36 | **20.68** |
|  | gpt2-med | 33.83 | 52.70 | 19.73 |
|  | scratch | 16.05 | 22.54 | 7.63 |

|  |  | METEOR | USE | BLEU |
|---|---|---|---|---|
| t=0.6 | jam | 33.41 | 51.08 | 20.42 |
|  | jam-so | 34.11 | **52.75** | 20.16 |
|  | jam-sojm | **34.73** | 52.27 | **21.12** |
|  | gpt2-med | 33.99 | 52.66 | 20.19 |
|  | scratch | 15.55 | 21.43 | 7.23 |

The two tables above show the performance of the fine-tuned models according to three metrics recommended for evaluating code summarization techniques [15]. We show results for the full holdout set as well as when removing methods that match the threshold t=0.6 for jm52m. We make the following observations:

1. The jam model and its two "brothers" achieve performance exceeding the gpt2-medium baseline, which was trained on a larger but proprietary dataset.
2. The jam-so model slightly outperforms jam for this problem, which is not surprising because so13m contains much more natural language data and code summarization is a natural language generation task.
3. The scratch model performs poorly, which is not surprising given the large model and small (170k) dataset.
4. The results change slightly at a different threshold $t$.

## 5 DISCUSSION / CONCLUSION

This tool demonstration advances the community by providing a language model and supporting tools for research problems involving Java source code. We intend our model especially for researchers working at a relatively fine level of granularity: Java methods, statements, and variables. Our model is a "one stop shop" for experiments involving fine-tuning, as we provide domain-specific pretraining datasets, multiple model configurations, fine-tuning tool support, and a deduplication toolkit to help ensure scientific integrity. We provide a holdout set of over 8k Java methods, as well as support for deduplicating one's own test set. We even provide traceability of each Java method in jm52m to its file and project of origin.

We have demonstrated our tool in an application for the problem of code summarization. We show how jam is able to outperform the most-similar proprietary baseline (gpt2-med) on this task under identical conditions. We note that the score we report are higher than those reported in recent code summarization papers for the same dataset [2, 3]. It is likely that these results would be surpassed with a larger model pretrained on more data. But, our finding is in an environment where we can control every experimental variable from pretraining data, to model architecture, to training parameters, and fine-tuning details. We are able to exercise this level of control at low costs: all experiments can be reproduced from scratch on a workstation with only a single 16GB GPU.

Our idea is that jam will fill the role of a "mouse model" for fine-tuning experiments involving Java methods, statements, and variables. Researchers with ideas about how to improve language models for some task can use this tool as a highly-controlled testbed to create a proof-of-concept. A researcher can try many permutations of the idea at low cost and then scale up the idea to a larger model where the researcher has less control of variables and much higher expense.

The doorway to our tool is our website:

**https://github.com/apcl-research/jam**

At that link, readers will find the following key components:

**datasets** via Huggingface repositories.

**model releases** via Huggingface repositories.

**dedup/finetune toolkits** via the Github repository.

**application demo** via the Github repository.

**instruction manual** via the Github repository.

## ACKNOWLEDGEMENT

We thank Andrej Karpathy and Daniel Grittner for their work providing the NanoGPT and NanoGPT-LoRA code. This work is supported in part by NSF CCF-2100035 and CCF-2211428. Any opinions, findings, and conclusions expressed herein are the authors and do not necessarily reflect those of the sponsors.

# 6 APPENDIX

In this Appendix we provide detailed instructions for replication of our results. Please download the source code from our Github repository linked in Section 5 and follow following steps:

- If you only want to finetune one of our pre-trained models, refer to subsections 6.1, 6.2, and 6.4.
- If you only want to deduplicate your dataset, refer to subsection 6.3.
- If you want to re-train a model using our processed and tokenized dataset, refer to subsection 6.7.
- if you want to scratch-train, by reprocessing the dataset, refer to subsections 6.6 and then 6.7.

We also present a video walk-through of our tool here:

https://www.youtube.com/watch?v=WP7ya17uYcY

## 6.1 Model Checkpoints

The first step is to download our pre-trained model checkpoints. They are downloaded as directories that contain weights and checkpoints needed for finetuning or retraining. We have made the following model checkpoints publicly available, each trained to one epoch, i.e., roughly 300K iterations.

**jam** - https://huggingface.co/apcl/jam
**jam-so** - https://huggingface.co/apcl/jam_so

We also provide checkpoint for the model trained twice, for one epoch on each dataset. Note, the learning rate and decay were reset between each epoch.

**jam-sojm** - https://huggingface.co/apcl/jam_sojm

Any of these model checkpoints can be downloaded using the following script:

```
python3 download.py --repo_id=apcl/jam --
    repo_type=model
```

This python script can be used with several tags to download all or specific data items from a respository:

- **--repo_id** specifies the name of repository, which for models are apcl/jam, apcl/jam_so, or apcl/jam_sojm.
- **--filename** specifies the name of the specific file that you want to download from the repository (optional).
- **--local_dir** specifies the name of your local output directory. The default value for this flag is the "data" directory that is populated from our github repository.
- **--repo_type** specifies the type of repository that hosts the file, i.e., "model" for these repositories.

If you do not want to use our pre-trained checkpoints, you may train the model from scratch using instructions in Section 6.3.

## 6.2 Finetuning

The next step is finetuning one of the model checkpoints. We make the training and test data to finetune our model checkpoints for source code summarization available at:
https://huggingface.co/datasets/apcl/funcom-java-long
To fine-tune the apcl/jam checkpoint for source code summarization as described in Section 3.6, please run the following commands:

```
cp -r jam jam_ft
torchrun --standalone --nproc_per_node=XX
    train.py config/finetune_funcom.py
    --out_dir=jam_ft
```

Here, the config/finetune_funcom.py provides the configuration required for finetuning and –out_dir specifies the path to the pre-trained model checkpoints. We make a copy of the pre-trained weights because the script modifies the files in the –out_dir during finetuning.
The following commands can be used to fine-tune using the apcl/jam_so checkpoint:

```
cp -r jam_so jam_ft
torchrun --standalone --nproc_per_node=
    whatever train.py config/
    finetune_funcom.py --out_dir=jam_ft
```

## 6.3 Deduplication

The next step is to run the de-duplication tool, which is described in Section 6.3. To test for deduplication over the jm52m dataset, please run the following command:

```
python3 data/jam_jm52m/dedup_fctest.py
```

This script can be used with the following flags:

- **--test_filename** specifies the path to the the test file.
- **--lsh_dir** specifies the directory for LSH files.
- **--threshold** specifies the threshold to which the test function and training function can be similar before they are considered duplicates. We recommend a threshold for 0.70 over our dataset.
- **--dedup_outfile** specifies the output file, where each entry is the function id tab-limited with a **set** duplicate functions id in the current **part** of the training set. More on parts below.
- **--fundats_file** specifies the name of the raw code file that is a dictionary for raw function code with key = function id and value = raw code. This file can be downloaded for our dataset, instructions for which are in Section .

The deduplication process as described in Section 6.3 relies on system memory. We provide these additional flags to decrease this computational load by dividing the training data into 50 parts. You may iterate through these parts, as memory allows using these flags:

- **--partstart** specifies the starting part number of the dataset, with a minimum value of 0.
- **--partend** specifies the ending part number of the dataset, with a maximum value of 50.

To test for deduplication over the so13m dataset, please run:

```
python3 data/jam_so13m/
    dedup_stackoverflow.py
```

This script can be used with the following flags:

- **--stackoverflow_text_id_filename** specifies the path to the pickle file that is a list for apcl/so13m file names.

- **--fundats_file** specifies the path to a pickle file that is a dictionary for raw function code files, where key = function id and value = raw code.
- **--stackoverflow_text_filename** specifies the path to a pickle file that is a dictionary for apcl/so13m posts with key = post id and value = post.
- **--dedup_outfile** specifies the path to the output file, where each entry is the function id tab-separated with a **set** of post IDs that are duplicate in the current **part** of the training set.
- **--threshold** specifies the threshold to which the test function and a post can be similar before they are considered duplicates.
- **--test_filename** file name of your test file
- **--lsh_outdir** directory for lsh files
- **--partstart** specifies the starting part number of the dataset, with a minimum value of 0.
- **--partend** specifies the starting part number of the dataset, with a minimum value of 100.

Note, it is possible that each test ID may have several entries in the output file because the tool works in parts to limit system memory requirements.

### 6.4 Test Set

The penultimate step is to extract the test set. To download and extract the test set, run the following command:

```
python3 download_extract_file.py
```

This script can be used with the following flags:

- **--repo_id** the id of repository that you want to download files
- **--local_dir** directory that you want to put your files
- **--filename** name of the file that you want to download

### 6.5 Inference

The final step is to run inference and predict summaries using the source code from the test set, using the following command:

```
python sample_funcom.py --out_dir=outdir
```

Note, the directory specified with the `out_dir` tag must be the directory where the final model weights to be used are saved. This script generates a prediction directory with a text file, where each line is the function id tab-separated by the predicted summary sequence.

### 6.6 Scratch-Compile Dataset

To scratch-compile our dataset, use the following command:

```
python3 download.py --repo_id=apcl/jm52m
    --filename=*.pkl --repo_type=
    dataset
```

The candidates for `repo_id` values are `apcl/jm52m` and `apcl/so13m` respectively. The script above downloads the raw data as a pickle file `fundats-j1.pkl`. Also, a list of function ids in our code summarization test set `q90testfids.pkl` that we exclude from the training set. Now using these files, a training dataset can be generated using the following command:

```
python3 data/jam_jm52m/prepare_fc_raw.py
    --num-proc=4 --q90testfids-file=
    q90testfids.pkl --fundats-file=
    fundats-j1.pkl
```

Here, `q90testfids.pkl` is a list of function ids from our test set that we exclude from the training set. This script will generate the train,val, and test bins required to retrain the model as described in the next subsection.

### 6.7 Re-Training Instructions

We also provide instructions for re-training our models if the pre-trained checkpoints are not desirable. The first step for scratch training is to download the required datasets. We provide public access to both datasets described in Section 3.3 at:

**jm52m** - https://huggingface.co/datasets/apcl/jm52m
**so13m** - https://huggingface.co/datasets/apcl/so13m

We also provide a script in our github repo to download these datasets using the follow command:

```
python3 download.py --repo_id=apcl/jm52m
    --filename=train.bin --repo_type=
    dataset
```

The candidates for `repo_id` values are `apcl/jm52m` and `apcl/so13m` respectively. Note, the `repo_type` is "dataset" to download datasets from our repositories. Note, without the use `filename` tag, the script will download the entire dataset hub, which includes roughly 50 GigaBytes of LSH MiniHash files for deduplication as described in Section 6.3.Please refer to Section 6.1 for a list of flags that can be used with this script.

Next, we train the model from scratch using the following torchrun command:

```
torchrun --rdzv-backend=c10d --rdzv-
    endpoint=localhost:0 --nnodes=1 --
    nproc-per-node=1 train.py config/
    train_funcom_raw.py --out_dir=
    jam350m_jm
```

Note, these configuration files define the dataset that is used to train the model:

- **train_funcom_raw.py** to train the model over the jm52m dataset
- **train_stackoverflow.py** to train the model over the so13m dataset

Please note, when using torchrun, the port number for rdzv-endpoint may be changed for multiple instances on the same machine using the following document.
https://pytorch.org/docs/stable/elastic/run.html
Otherwise, two different training instances may update the same model weights.

### 6.8 Hardware

We recommend a GPU with an architecture comparable to the NVidia Ampere [27] or newer, because the "bfloat16" format is essential for efficient computation with our scripts. For GPUs older

Chia-Yi Su, Aakash Bansal, Vijayanta Jain, Sepideh Ghanavati, and Collin McMillan

than that, "float32" format may be used. However, the VRAM requirements are higher using that format and computations are considerably slower.

Note, a workstation with NVidia A4000 GPU can be assembled for under US$1500 as discussed in Section 3.1.